# Unravelling the Asymmetric Volatility Puzzle:

## A Novel Explanation of Volatility Through Anchoring

**MIHÁLY ORMOS**

Department of Finance, Budapest University of Technology and Economics
Magyar tudosok krt. 2., 1117 Budapest, Hungary
Phone: +36 1 4634220; Fax: +36 1 4632745
ormos@finance.bme.hu

**DUSAN TIMOTITY**

Department of Finance, Budapest University of Technology and Economics
Magyar tudosok krt. 2., 1117 Budapest, Hungary
Phone: +36 1 4634220; Fax: +36 1 4631083
timotity@finance.bme.hu

**Abstract**

This paper discusses a novel explanation for asymmetric volatility based on the anchoring behavioral pattern. Anchoring as a heuristic bias causes investors focusing on recent price changes and price levels, which two lead to a belief in continuing trend and mean-reversion respectively. The empirical results support our theoretical explanation through an analysis of large price fluctuations in the S&P 500 and the resulting effects on implied and realized volatility. These results indicate that asymmetry (a negative relationship) between shocks and volatility in the subsequent period indeed exists. Moreover, contrary to previous research, our empirical tests also suggest that implied volatility is not simply an upward biased predictor of future deviation compensating for the variance of the volatility but rather, due to investors' systematic anchoring to losses and gains in their volatility forecasts, it is a co-integrated yet asymmetric over/under estimated financial instrument. We also provide results indicating that the medium-term implied volatility (measured by the VIX Index) is an unbiased though inefficient estimation of realized volatility, while in contrast, the short-term volatility (measured by the recently introduced VXST Index representing the 9-day implied volatility) is also unbiased and yet efficient.

**Acknowledgements:** We are grateful to the conference participants at the Workshop on Behavioural Economics and Industrial Organization at Corvinus University, 2014. We would like to gratefully acknowledge the valuable comments and suggestions of the anonymous referee that contribute to a substantially improved paper. Mihály Ormos acknowledges the support by the János Bolyai Research Scholarship of the Hugarian Academy of Sciences. Dusán Timotity acknowledges the support by the Fundation of Pallas Athéné Domus Scientiae.

# Unravelling the Asymmetric Volatility Puzzle
## A Novel Explanation of Volatility Through Anchoring

**Abstract**

This paper discusses a novel explanation for asymmetric volatility based on the anchoring behavioral pattern. Anchoring as a heuristic bias causes investors focusing on recent price changes and price levels, which two lead to a belief in continuing trend and mean-reversion respectively. The empirical results support our theoretical explanation through an analysis of large price fluctuations in the S&P 500 and the resulting effects on implied and realized volatility. These results indicate that asymmetry (a negative relationship) between shocks and volatility in the subsequent period indeed exists. Moreover, contrary to previous research, our empirical tests also suggest that implied volatility is not simply an upward biased predictor of future deviation compensating for the variance of the volatility but rather, due to investors' systematic anchoring to losses and gains in their volatility forecasts, it is a co-integrated yet asymmetric over/under estimated financial instrument. We also provide results indicating that the medium-term implied volatility (measured by the VIX Index) is an unbiased though inefficient estimation of realized volatility, while in contrast, the short-term volatility (measured by the recently introduced VXST Index representing the 9-day implied volatility) is also unbiased and yet efficient.

## 1. Introduction

Despite volatility forecasting models having undergone a long evolution, novel regressions still have difficulty estimating the deviation of future returns. Currently, the most common techniques including ARCH-type models (Lamoureux and Lastrapes, 1993; Blair et al., 2001) and model-free volatility forecasts (Carr and Madan, 1998; Demeterfi et al., 1999; Britten-Jones and Neuberger, 2000; Carr and Wu, 2009) fail to explain the effects of jumps and non-continuities in the price of the underlying asset. However, Taylor et al. (2010) argue that methods based on implied volatility ("IV") are likely to outperform the aforementioned methods if the prediction horizon extends until the expiry date of the options. Becker et al. (2009) also conclude that IV itself contains more information than individual model-based forecasts. For short-term volatility, our paper supports the ideas of Taylor et al. (2010) and Becker et al. (2009) as we measure that IV subsumes the information contained in historical realized volatility ("RV"). Notwithstanding, in the long run we do find that although being co-integrated with RV, IV is an unbiased yet inefficient estimation of the future deviation of returns.

Standard volatility autoregression models imply that massive changes in the price of the underlying asset should result in increased RV over the next period, however, our findings show that volatility increases only in case of losses. This phenomenon is known as asymmetric volatility. Despite the wide amount of literature devoted to asymmetric volatility (Black, 1976; Christie, 1982; and Schwert, 1989), which we discuss below in detail, no adequate explanation for this has been given yet. In this paper, we propose a novel explanation for this phenomenon based on the anchoring heuristic (Kahneman and Tversky, 1974 and, 1979).

According to Fleming (1998) and Andersen et al. (2003), by applying a model that corrects the IV's upward biased forecast, implied volatility has a greater predictive power. In contrast, we find that analyzing the relationship in the context of asymmetric volatility, their results only hold in the case of price declines in the underlying asset. Since white noise effects cancel out, on average, these two types of volatility reflect the same level of standard deviation (as measured by our co-integration tests and by the fractional co-integration results of Bandi and Perron (2006)). Moreover, the IV of a given period is conditional to the filtration at the beginning of the interval (*ex ante*) whereas the RV is measured at the end (*ex post*). Therefore, through an analysis of the relation between the two, effects having a different influence at the beginning and during a given period can be separated. Furthermore, in our proposed reasoning for this asymmetric volatility the anchoring exactly follows this type of effect, thereby fading out over time. Hence, we may bolster our theory with empirical support by describing the relation of RV and IV through regressions and event studies.

We also provide robustness tests in two ways: firstly, we test whether anchoring is also present in short-term analysis and secondly, we examine alternative intervals of the return distribution; in other words we test whether the asymmetry is present only in case of extreme circumstances or generally present throughout.

This paper is organized as follows: Section 2 discusses our anchoring-based explanation for the asymmetric effect. Section 3 provides a review of the data and methodology used. In Section 4 our empirical results are presented in detail and, lastly, in Section 5 we offer a brief conclusion of the main contribution of our research.

## 2. Anchoring and asymmetric volatility

Debate has raged over the last few decades as to which factor is responsible for the asymmetric volatility effect (i.e. are stock returns negatively correlated with the volatility of the next period?). Bae et al. (2007) argue that there are currently two main explanations for the asymmetric volatility puzzle. The first is the leverage effect noted by Black (1976), Christie (1982) and Schwert (1989). These authors find, if rather simply, that if the value of an equity drops the firm becomes more leveraged thereby causing the volatility of equity returns to rise. All three authors conclude that, although volatility is indeed an increasing function of financial leverage, the effect by itself is not sufficient to account for the observed negative correlation. The second explanation is the volatility feedback hypothesis, which states that in cases of unexpected increases in volatility, future expected volatility rises and thus the required return of the stock also increases and results in an immediate negative impact on current stock price. Although numerous studies have found evidence in support of both explanations (Pindyck, 1984; Kim et al., 2004), recent studies still provide results that frequently contradict each other; Bollerslev et al. (2006) finds that analysis of high-frequency data indicates no significant volatility feedback, while Bekaert and Wu (2000) conclude that the leverage effect is insignificant. We aim to introduce an alternative explanation for this conundrum.

Since the milestone paper of Kahneman and Tversky (1974), much research has provided support for the existence of anchoring (Ariely (2003), De Bondt (1993) and Shefrin (2002)). De Bondt (1993) suggests that anchoring consists of two separate effects.

On the one hand, price anchoring (De Bondt (1993)) leads people to rely heavily on average past price levels, and therefore an unexpected exogenous shock in the price could imply a belief in mean-reversion and lead to, in the short term, under/overpricing without any significant effect on fundamental value (see also De Bondt and Thaler, 1985; Daniel et al., 1998; Barberis et al., 1998; Chopra et al., 1992).

On the other hand, De Bondt (1993) underlines that, in addition to recent price levels, past returns (i.e. previous trends) are also used as anchors in investors expectations; therefore, linear extrapolation of bullish and bearish trends are also present in estimations of future prices. This latter might be attributed either to the representativeness heuristic, which would imply that investors interpret the recent trend as a representative pattern for the whole time-series, or to availability heuristic implying that investors pay much more attention to recent information. However, as in De Bondt (1993) we argue that, instead of the alternative heuristics, both of the anchors (i.e. the price level and its recent change) are in focus, and therefore, they together form the future expectations. According to his results, in contrast to alternative explanations implying that the recent trend is believed to continue (e.g. the representativness heuristic), the subjective probability given to the continuing trend is much higher than that of a mean-reverting process following bull markets, whereas in case of bear markets the difference is insignificant. Therefore, due to the asymmetric distribution of trend followers and contrarians, previous gains yield more identical expectations of the future price resulting in lower anticipated volatility than losses that equally include estimates of both mean-reverting and trend continuing predictions. In other words, in case of bull markets the effect of the price change anchor (i.e. continuing trend) dominates over price level effect (i.e. mean-reversion), whereas, in case of bear markets both anchors equally contribute to the prediction of future prices hence increasing uncertainity, which latter leads to a higher uncertainty of estimated future prices.

Having formed the expected (i.e. implied) volatility, the last step in the effect of previous anchors on future risk consists of the relationship between the estimated and realized volatility. Accorrding to the law of one price forward and option prices should converge to current spot prices and vice versa (Cox et al. (1981), Fama (1984)). Therefore, through the convergence of prices, IV affects RV and results in a positive linear relationship. Hence, there is a respective increase or decrease in IV and RV following negative or positive returns. Despite this relationship, anchoring fades away over time, and therefore as time progresses, its effect (i.e. the effect of a previous shock and a price level) on RV should be less significant as new information comes to light. Hence, by comparing the effects of a given shock (i.e. the anchor) on subsequent implied and realized volatility, which two are cointegrated (and therefore are driven by similar processes in the long-term), the fading property of the previous anchors can be tested.

Another explanation based on anchoring could be attributed to the change of investors' risk-aversion. One of the most successful theories in describing economic choice, the Prospect Theory (Kahneman and Tversky, 1979) suggests that utility perception is reference-dependent and is a result of *changes* in wealth rather than its total value. Using the rational expectations of outcomes as this reference point as suggested by Koszegi and Rabin (2006) the following effects emerge. Loss-aversion induces aggregating returns without realizing the loss; therefore, the previous abnormal return is used as a quasi-anchor (i.e. its negative value is included in the required and thus the expected return). Therefore, the hope of breaking-even following losses increases the attractiveness of riskier assets and thus decreases the perceived risk-aversion, which results in higher volatility following losses and lower volatility after gains.

However; the problem with changing risk-aversion is the ambiguity of the existing experimental results. Thaler and Johnson (1990) confirm the existence of a positive or negative relationship between previous abnormal returns and risk-aversion for choices where breaking-even is existent or absent respectively. They argue that the choice between risk seeking and risk averse behavior depends on the frame in the case of both positive and negative exogeneous price shocks: in their experiment they find that the certain behavior depends on mental accounts; they conclude that it cannot be identified which effect (that of the hope of breaking-even or the "gambling with the house money") dominates in a certain situation. Contrary, Fama (1998) argues that apparent overreaction of stock prices to information is about as common as underreaction. In this context frames suggest that decrease and increase in risk aversion exhibit uniform distribution, therefore, no significant pattern of any of them should exist at an aggregate level. In our paper we propose an explanation, the anchoring, that gives unambiguous results for investors' behavior in the case of exogeneous price shocks compared to the explanation of mental accounts and changes in risk aversion. We show that following a decrease in price risk-seeking, while after a price jump risk-averse behavior can be observed, which behavior is in line with anchoring and cause asymmetry in volatility.

Although any of the previous theories may contribute to asymmetric volatility, due to the absence of individually measured panel data, only aggregate effects are analyzed in this paper and individual factors such as change in risk-aversion cannot be separated.

After describing data and methodology, we go on to provide empirical support for the process mentioned above: first, we show that IV and RV are co-integrated (and hence their difference can be analyzed); second, we discuss our results, which indicate that the anchoring effect is, in fact, present.

## 3. Data and methodology

### 3.1 Data

We use historical data for realized and implied volatility. First, we calculate RV using the daily historical returns of the S&P 500 index for the period January 2, 1990 to August 07, 2014. To capture IV, we use the daily closing prices of two volatility indices provided by the Chicago Board of Options Exchange ("CBOE"). For long-term analysis we apply the VIX Index created to measure the market estimate of future S&P 500 volatility by monitoring put and call options with maturities close to 30 days. The VIX, with its new calculation method derived from the forward prices, reflects the annualized estimated volatility of S&P 500 over the next month (CBOE, 2009). The longevity of the period used as our main sample (24 years) covers significant upward and downward trend periods and thus includes a much higher number of observations than those applied in most studies on implied volatility. Moreover, a significant amount of non-overlapping data can also be used to make the comparison between monthly IV and RV.

For short-term analysis we use the, recently introduced, VXST Index which measures estimated volatility over the next 9 days in the same way as the VIX does (CBOE, 2014). CBOE has only been providing historical data for this instrument since January 3, 2011 hence, we use a mirror period for S&P 500 returns for the short-term comparison of IV and RV.

*3.2 Methodology*

Primarily we focus on the effects of massive price changes; thus, we investigate changes in the VIX during periods following abnormally high or low returns. As the VIX reflects the annualized estimated volatility over the next month, to facilitate a comparison, we define the daily realized volatility over the same interval and annualize it according to the method already used in the VIX calculations (CBOE, 2009). Analytically this means that

$$RV_t = 100\sqrt{\sum_{k=t+1}^{t+22} \frac{(r_k - \bar{r}_{t+1,t+22})^2}{22} 365} \qquad (1)$$

where $RV_t$ indicates the annualized, realized volatility in the month following the $t^{th}$ day in percentages, $r_k$ represents the daily log return of the investigated portfolio and $\bar{r}_{t+1,t+22}$ stands for the next month average daily log return. This process allows us to compare the current volatility implied by options expiring next month with the realized volatility in the next month and thus, determine the effectiveness of the estimation by implied volatility. We use non-overlapping data, hence every 22$^{nd}$ data point of RV and IV is compared.

We apply the same technique to describe short-term realized volatility, however, according to the 9-day volatility measured by VXST, the next 6 daily log returns are used to calculate realized volatility and in this case every 6$^{th}$ data point is analyzed.

Although one may think that the change of the VIX or the VXST over the preceding period should not necessarily reflect the exact change of the realized volatility over the next period since new information may affect the future deviation of returns, the random individual effects should, on average, cancel themselves out. We are, therefore, able to make this comparison as proven in the results of Section 4.1.

One of the main purposes of our research is to examine how massive price increases (price jumps) and massive price decreases (price falls) affect volatility estimated for, and realized in, the following month. Henceforth, by massive price changes we mean 22 day (or 6 day) returns below the 10$^{th}$ or above the 90$^{th}$ percentile of the distribution. In Section 4.3 we present a robustness test for our monthly (22 day) results, where we classify returns based on being above and below zero instead of using the percentiles mentioned above. In our other robustness test, where we analyze the short-term effect, the same classification of groups is used, since a lack of input data for VXST results in the number of observations being too low to draw any meaningful conclusion for extreme changes only.

In order to compare the effects of massive price changes on volatility measures we estimate linear regression models. Our regressions contain an indicator function that is equal to one if the previous return is a price fall (or below zero) and zero otherwise. By controlling other variables such as the previous return itself, we can test the significance of the asymmetry in case of both RV and IV.

Since our proposed reason for the asymmetry should be more significant in short-term analysis, we measure whether its effect on IV, which is measured at the time of the event, is stronger than the effect on RV, which is measured later in time.

## 4. Empirical results

### *4.1 The relationship between realized and implied volatility*

As we have argued in Section 2, implied volatility changes are comparable to realized ones. In order to prove this statement we firstly present Figure 1 which shows the evolution of the two processes for the observed period (1990-2013). Intuitively, one may see that calculated realized volatility (dashed) is fluctuating around the implied volatility (continuous), hence, individual effects (e.g. information revealed in the following month) seem to cancel each other out.

**Please insert Figure 1 about here**

However, in order to rigorously validate our statement we also test whether the two processes are convergent - is their difference stationary? In order to examine stationarity, we test the co-integration criteria using regressions with and without the intercept. The p-values of our results presented on Table 1 indicate that the intercept is not materially significant. One of the two main measures used in goodness-of-fit tests, the Akaike information criterion, suggests that the intercept model is a more suitable fit for the data. However, based on the Bayesian information criterion the no-intercept model is the better. The Student-t test for the residual of the no-intercept model indicates no significant mean. However, IV in the linear regression is very significant (p-value less than $2*10^{-16}$) and based on the Augmented Dickey-Fuller test no unit root is present. Therefore, the residual is stationary around zero, thus RV and IV are co-integrated both in short and long term and are described with the linear regression of $RV_t = 0.9474 \cdot IV_t + e_t$ and $RV_t = 0.8595 \cdot IV_t + e_t$ respectively. The summary of our co-integration results is presented in Table 1.

**Please insert Table 1 about here**

Apart from testing the co-integration we also test whether the model residual is a white noise process. Hence, according to Prabhala et al. (1998), our regression can be used to decide whether implied volatility is an unbiased instrument if the intercept is zero, and also whether it is efficient if IV subsumes all the information about historical RV. Our results, presented in Table 1, indicate that based on Student-t test IV is unbiased. However, one can see from the Box-Pierce test that autocorrelation is significant in the long term: we measure an AR(1) process for the residual. In other words, the monthly IV (VIX) is an unbiased though inefficient financial instrument, while its short term version (VXST) is also unbiased and yet efficient.

*4.2. Anchoring in implied and realized volatility*

Numerous studies related to the estimation of future volatility have shown that implied volatility is a more efficient measure than purely simple historical data since it contains incremental information beyond model-based forecasting (Christensen and Prabhala, 1998; Wu et al., 2002). As provided in Section 4.1, we measure that this is only valid in case of a short-term analysis. However, the co-integration criterion, which is valid in every case, is sufficient to make a comparison between IV and RV. Hence, for comparative purposes, in Table 2 we present the regression results of extreme price changes.

**Please insert Table 2 about here**

In the regressions presented we use the indicator function of

$$I_t = \begin{cases} 1 \; if \; r_t < 0 \\ 0 \; otherwise \end{cases}. \quad (2)$$

As a control variable for the indicator function our linear models apply previous period return, thus we can separate the asymmetric effect of return shocks. The results presented in Table 2 indicate two main findings relevant to our paper. First, the indicator function is negative and very significant in both cases, which provides support for: (i) previous shocks do have a significant effect on volatility in long term analysis and (ii) this effect is asymmetric for gains and losses. The former indicates that recent information (whether it is fundamentally relevant or not) plays an important role in driving future expectations. This phenomenon could be attributed to the representativness heuristic as well, however, the latter finding (i.e. the asymmetry) is a property specific to anchoring as discussed above. Here, seemingly irrelevant factors (i.e. the two anchors of the previous return and price level in this case) affect the estimations for future prices in an asymmetric way, that is, the bias in investors' predictions (i.e. the beliefs of continuing and mean-reverting trends) is different in the cases of previous gains and losses, which supports our theoretical explanation consisting of two distinct effects. Our second main finding indicated by Table 2 is that the p-value of the asymmetric factor is much lower for IV implying that anchoring is more significant; in other words the asymmetry in the previous period shock affects much more significantly the *ex ante* estimated implied volatility where the effect itself is not subject to time decay. This result also supports our theory of the asymmetric effect fading away, since estimation for implied volatility is made well before (*ex ante*) the realization of true volatility (*ex post*), therefore, the well-known time-decaying characteristic of anchoring is present. The coefficient of the previous period indicator function and return could also be interpreted as the anchoring index (AI) used by Jacowitz and Kahneman (1995) where the biased estimation is compared to the prediction without anchoring.

Moreover, our regression results are in line with recent findings of Kliger and Kudryavtsev (2013). They investigate the effects of market analysts' recommendations on VIX changes and found that positive (negative) excess returns following recommendation upgrades (downgrades) are stronger when accompanied by daily VIX decreases (increases). Their findings also support our linear estimation results, since, in addition to the negative coefficient of the market return in the RV regression, its coefficient is negative (although much less significant) in the IV regression as well. Hence, both RV and IV are negatively correlated with the market return; hence, due to the average co-movement of individual assets, the change in VIX may provide positive or negative excess return at individual level.

We provide further support for anchoring theory by event studies around extreme price changes. We use a time frame of (t-1; t+2), where cumulative return above the t-1 volatility level is measured. Our results are shown in Figure 2 where the cumulative change of RV and IV is represented by the dashed and continuous curves respectively.

**Please insert Figure 2 about here**

The graphs indicate two main findings: first, in the long term, volatility seems to be persistent and following price jumps and falls RV and IV has a measurable cumulative return above and below the t-1 reference point at t+1. Second, only price jumps seem to cause asymmetric realized volatility in the following period; subsequent to price falls RV does not change from t-1 to t and IV adjusts to elevated levels of risk. Although, both graphs indicate that at the end of the time frame RV and IV have approximately the same cumulative return (as suggested by our co-integration tests) to highlight this and shed light on their dissimilarities we present Figure 3. Here, the existence of co-integration is confirmed by the mean-reverting process to zero and the p-values, although, in the case of price jumps the convergence seems to be slower. The difference between IV and RV is decreasing in significance, that is, the effect of anchoring to a shock diminishes over time.

**Please insert Figure 3 about here**

Our subsequent analysis of the effect of short-term price changes on long-term volatility also confirms our anchoring based theory. As mentioned above, anchoring to price falls (or jumps) increases (or decreases) IV first, and then, if anchoring is still present, RV is also affected. Thus, by testing short-term price changes (which fade away quickly in the long term), IV should exhibit asymmetry at a much more significant level. In other words, the effect of an anchor from the past decays over time, therefore, the difference between the *ex-ante* measured IV and *ex-post* measured realized volatility should be less significant for increasing length of period. Table 3 shows the effect of daily price changes on IV and RV of the following month. Here, we regress the non-overlapping sample of implied and realized volatility on their previous value (IV and RV in the t-1$^{th}$ month), on the S&P 500 return over the current t$^{th}$ month, on the S&P 500 return of the last day of the previous month and on the indicator function taking on non-zero values if the latter variable is negative. The results confirm that in the RV regression the last day's return is not significant, moreover, the p-value of the indicator function is much higher supporting the theory that short-term anchoring almost entirely fades away in the long term.

**Please insert Table 3 about here**

*4.3. Robustness tests*

In order to test the generality of our results we provide further support in the two following ways. First, we check whether the asymmetry and the effect of anchoring is solely present in the case of extremities or whether they are applicable to the whole return distribution separated into positive and negative groups. Second, we test whether short-term volatility behaves in the same manner as its long-term equivalent (that is the asymmetric volatility effect is present and is fading away in time).

The results of the first test, presented in Table 4, indicate that there is an inverse relationship between negative previous returns and both realized and implied volatility. Positive returns have significant negative effect on future implied volatility and a marginal effect on RV, while negative returns increase both volatilities. These effects together yield a negative correlation between return and volatility corroborating the presence of asymmetry. Furthermore, the effect on IV is much stronger; hence, the time-related characteristic of anchoring can be recognized again.

**Please insert Table 4 about here**

Figure 4 and 5 show event studies using the aforementioned return classification. As suggested by the regression results, the graphs are very similar to the ones measured in extreme cases, however, after the initial drop or jump at *t* the difference immediately reverts to zero. This finding reinforces our co-integration results and the assumption that, on average, the volatility estimation seems to be unbiased.

**Please insert Figure 4 about here**

**Please insert Figure 5 about here**

Our second robustness test, shown in Table 5, also reconfirms the existence of asymmetry and the anchoring effect: the indicator function of the short-term regression is significant, while the p-value is lower for IV estimation as discussed for alternative samples mentioned above. As Figure 5 shows the significance of the difference between IV and RV drops immediately following the shock, therefore, the anchor does not have any further effect on subsequent estimates of volatility (time-decaying).

**Please insert Table 5 about here**

Short-term analysis, presented in Figures 6 and 7, yields almost the same results as when measured in the long term. Both graphs indicate that, as proposed by our theory, anchoring is stronger here. The asymmetric volatility caused by the shock in *t* does not fade away in *t+1*; both from *t-1* to *t* and from *t* to *t+1* the effect is present, although, the latter is much less significant. The difference between IV and RV reverts to zero, which again is in line with co-integration.

**Please insert Figure 6 about here**

**Please insert Figure 7 about here**

Due to a mismatch in dates and numbers of observations, the unclassified comparison of the regression results would lead to biased conclusions. Therefore, only IV and RV models which use exactly the same input data are compared. However, our general findings are that asymmetric volatility is significant in every case (yielding a volatility modifying component of anchoring) and IV is more sensitive to the indicator function than RV, which indicates that shocks affect more significantly the ex-ante measured volatility (time-decaying component). Hence, we conclude that anchoring seems to be a valid explanation for asymmetry in volatility.

## 5. Concluding remarks

Despite asymmetric volatility attracting a great deal of attention, there still seems to be no consensual solution to the puzzle. Our paper presents novel findings in two fields:

First, we contribute to existing literature on realized and implied volatility by studying their relationship in the context of asymmetric volatility and by analyzing the, recently introduced, short-term implied volatility indicator, the VXST.

Second, supported by our analysis, we propose a novel explanation for the asymmetry puzzle by applying the behavioral pattern known as anchoring.

Our regression tests of the relationship between RV and IV yields that, in contrast to the prior literature, the two processes are co-integrated both in long and short term. Furthermore, our event studies confirm the zero-reverting difference between the two both in extreme cases as well as in general.

We conclude, through an economic interpretation of the analysis, that IV is an unbiased estimator of RV. However, using the VIX we find significant autocorrelation between the residuals in the long term, therefore, our results indicate that only the short term estimation, using the VXST, is efficient.

Our empirical results also support our proposed theory for asymmetric volatility. We find that the asymmetric effect of price falls and jumps is less significant on RV (which is measured ex post over the following period) than their effect on IV (that is estimated ex ante at the beginning of the following period). Hence, we conclude that both the asymmetric effect on volatility and the fading property of anchoring is supported by empirical results. Moreover, our short term event studies suggest that, in contrast to the long term results, the asymmetric effect is slightly present in following period as well; thus, the time-related characteristic of our proposed explanation is valid.

Nevertheless, there are still many questions on the topic which require further research. Less is known about the transition of IV of options into the RV of spot prices, the effect of changes in investors' risk-aversion on the asymmetry or about the implementation of the asymmetric effect into derivative pricing. These problems remain to be solved in future studies.

# Figures

## Figure 1

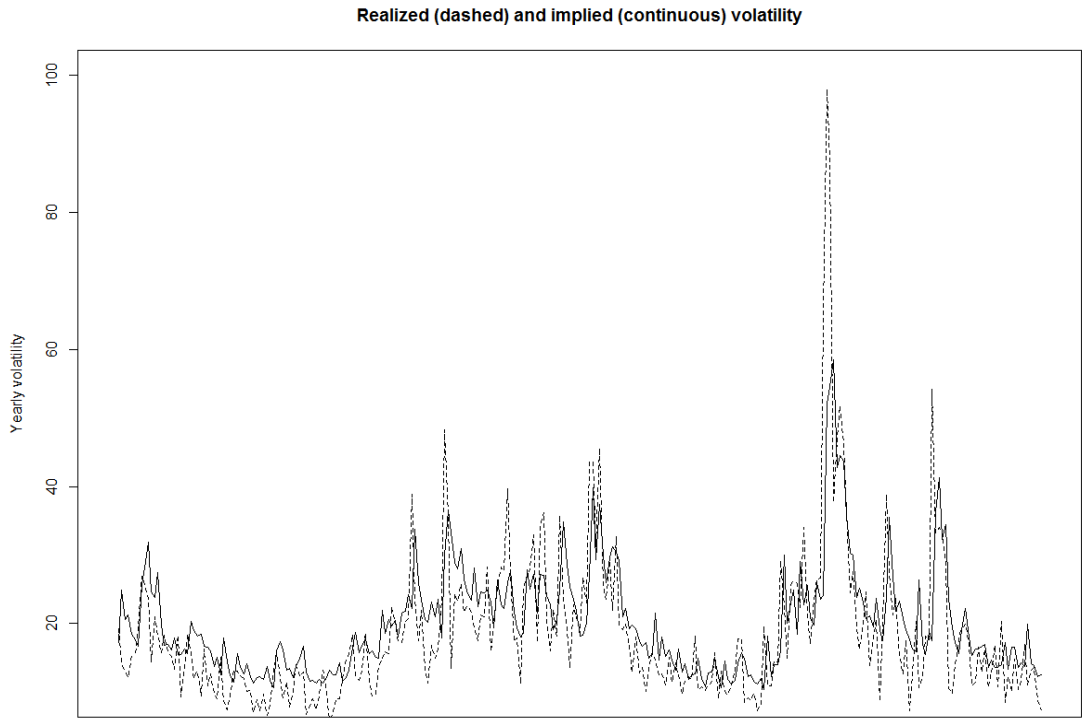

Notes: The horizontal and vertical axes represent the time for the period 1990/01/02-2014/08/07 and the volatility in percentage respectively. The continuous curve stands for the VIX (using new calculation method), while the dashed curve shows the annualized realized volatility of S&P500 in the next month

## Figure 2

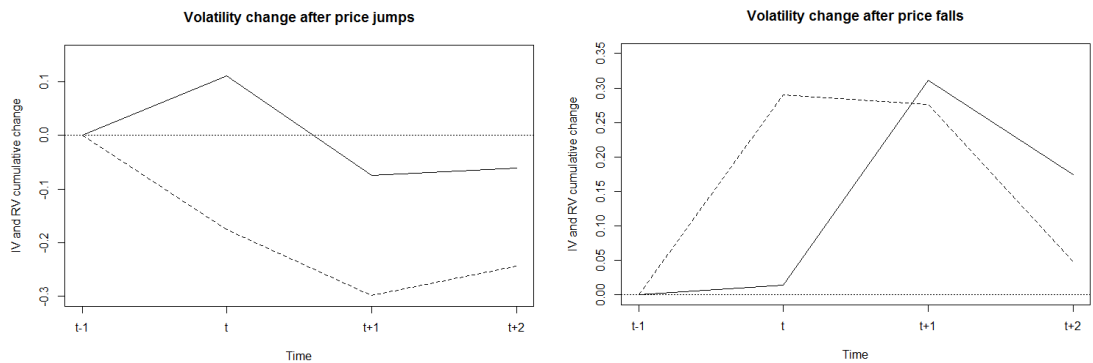

Notes: The horizontal and vertical axes represent the 22-day time steps and the cumulative return of volatility respectively. The continuous curve stands for the VIX cumulative return, while the dashed curve shows the change of realized volatility. The horizontal line represents the zero change.

# Figure 3

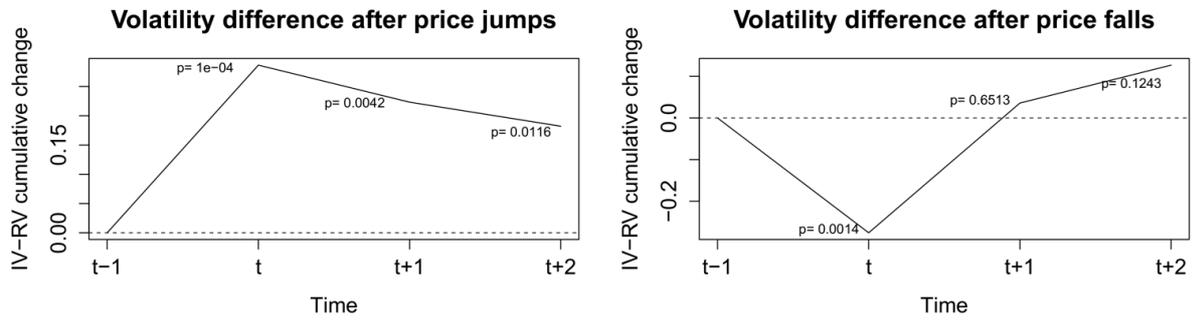

Notes: We define the difference as the cumulative return of IV predicted for month t less the cumulative return of RV measured over month t. The values next to the data points represent the p-values of the Student-t tests for zero mean.

# Figure 4

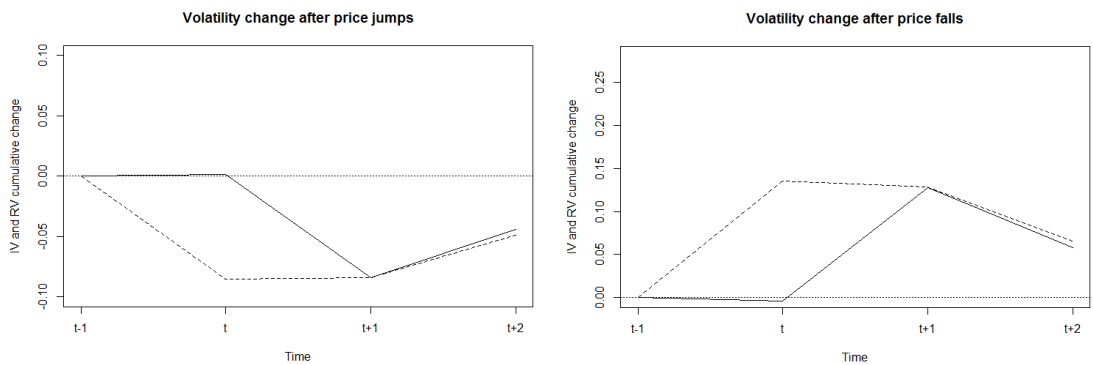

Notes: The horizontal and vertical axes represent the 22-day time steps and the cumulative return of volatility respectively. The continuous curve stands for the VIX cumulative return, while the dashed curve shows the change of realized volatility. The horizontal line represents the zero change.

# Figure 5

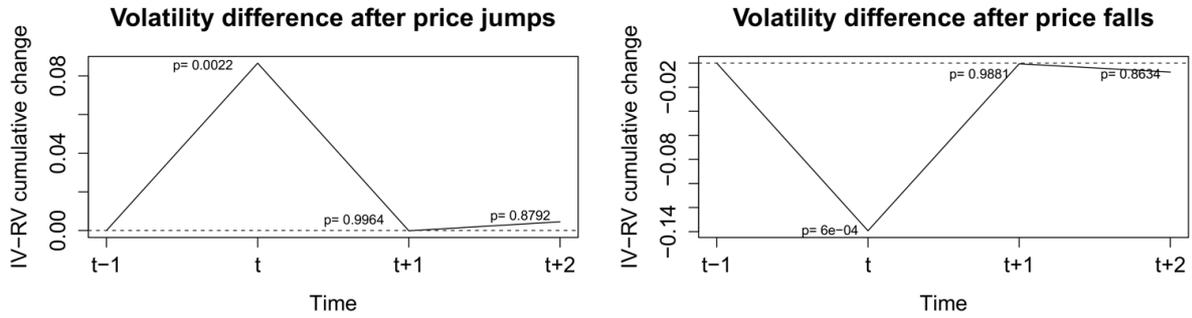

Notes: We define the difference as the cumulative return of IV predicted for month t less the cumulative return of RV measured over month t. The values next to the data points represent the p-values of the Student-t tests for zero mean.

# Figure 6

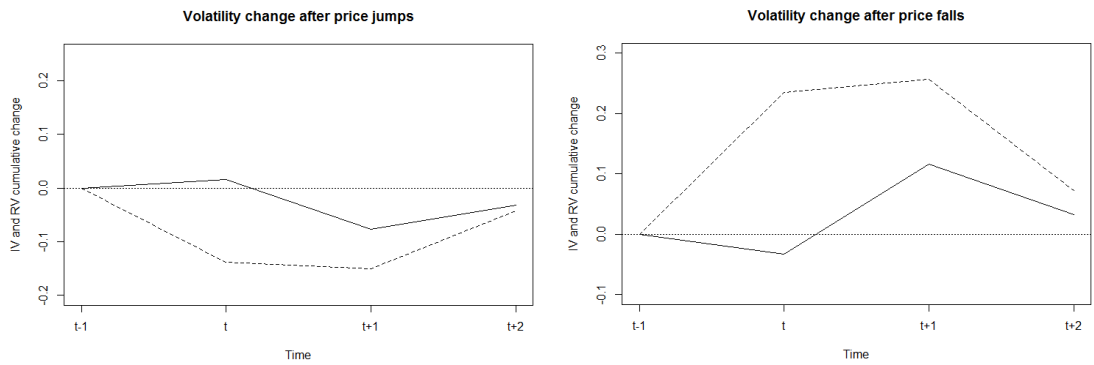

# Please insert Figure 7

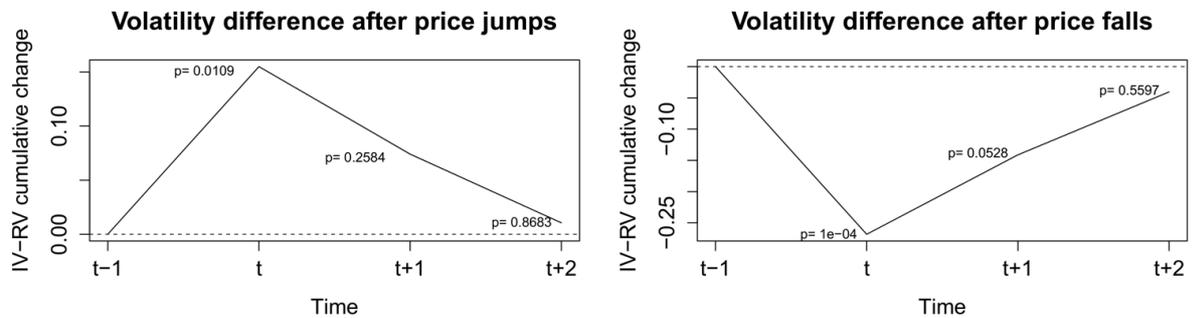

# Tables

## Table 1

| Co-integration analysis | | |
|---|---|---|
| Test | Short term | Long Term |
| **Student t-test of the residual** | 0.3936 | 0.2738 |
| **Box-Pierce test for autocorrelation in the residual** | 0.7285 | 1.79E-06 |
| **ADF test for unit root** | <0.01 | <0.01 |

Notes: The Box-Pierce test uses the lag of 1 to measure autocorrelation in the residual. The ADF test uses the lag order of 6 to test unit root and stationarity.

## Table 2

| | Extreme case in long term | | | |
|---|---|---|---|---|
| | RV | | IV | |
| Variable | Coeff | p-value | Coeff | p-value |
| **Intercept** | 3.5416 | 0.3841 | 3.662 | 0.0581 |
| **S&P 500 return in t-1** | 69.6316 | 0.2012 | -16.3083 | 0.4265 |
| **S&P 500 return in t** | -83.3115 | 0.0002 | -9.9148 | 0.2117 |
| **Autoregression for t-1** | 0.49 | 9.31E-08 | 0.7398 | 0 |
| **Indicator function in t-1** | -252.501 | 0.0110 | -108.855 | 0.0032 |

Notes: The results represent the following linear regressions using OLS estimation.

$RV(t) = \mu + \beta_1 \cdot r(t-1) + \beta_2 \cdot r(t) + \beta_3 \cdot RV(t-1) + \beta_4 \cdot I(t-1) \cdot r(t-1)$ and

$IV(t) = \mu + \beta_1 \cdot r(t-1) + \beta_2 \cdot r(t) + \beta_3 \cdot IV(t-1) + \beta_4 \cdot I(t-1) \cdot r(t-1)$.

## Table 3

### Effect of short-term price changes on long term volatility

| Variable | RV | | IV | |
|---|---|---|---|---|
| | Coeff | p-value | Coeff | p-value |
| Intercept | 5.6837 | 1.67E-12 | 3.7161 | 2.81E-08 |
| S&P 500 last day return in t-1 | 45.1042 | 0.488 | 206.2695 | 9E-09 |
| S&P 500 return in t | -70.0355 | 6.76E-13 | 9.7117 | 0.0751 |
| Autoregression for t-1 | 0.6411 | 0 | 0.7044 | 0 |
| Indicator function in t-1 | -343.1011 | 0.0036 | -520.4814 | 2.37E-16 |

Notes: The results represent the following linear regressions using OLS estimation.

$RV(t) = \mu + \beta_1 * r_{last}(t-1) + \beta_2 * r(t) + \beta_3 * RV(t-1) + \beta_4 * I(t-1) * r_{last}(t-1)$ and

$IV(t) = \mu + \beta_1 * r_{last}(t-1) + \beta_2 * r(t) + \beta_3 * IV(t-1) + \beta_4 * I(t-1) * r_{last}(t-1)$.

## Table 4

### General case in long term

| Variable | RV | | IV | |
|---|---|---|---|---|
| | Coeff | p-value | Coeff | p-value |
| Intercept | 6.7485 | 2.93E-15 | 3.4377 | 3.50E-14 |
| S&P 500 return in t-1 | 10.7562 | 0.5480 | -18.16 | 0.0153 |
| S&P 500 return in t | -84.4618 | 0 | 0.4022 | 0.9074 |
| Autoregression for t-1 | 0.5586 | 0 | 0.759 | 0 |
| Indicator function in t-1 | -127.2808 | 7.58E-05 | -101.589 | 2.08E-15 |

Notes: The results represent the following linear regressions using OLS estimation.

$RV(t) = \mu + \beta_1 * r(t-1) + \beta_2 * r(t) + \beta_3 * RV(t-1) + \beta_4 * I(t-1) * r(t-1)$ and

$IV(t) = \mu + \beta_1 * r(t-1) + \beta_2 * r(t) + \beta_3 * IV(t-1) + \beta_4 * I(t-1) * r(t-1)$.

**Table 5**

| | General case in short term | | | |
|---|---|---|---|---|
| | RV | | IV | |
| Variable | Coeff | p-value | Coeff | p-value |
| **Intercept** | 5.7686 | 7.40E-09 | 4.0897 | 8.02E-10 |
| **S&P 500 return in t-1** | 6.5788 | 0.8805 | -97.6698 | 6.48E-05 |
| **S&P 500 return in t** | -154.838 | 1.01E-10 | -9.3671 | 0.4030 |
| **Autoregression for t-1** | 0.5391 | 3.14E-15 | 0.7338 | 0 |
| **Indicator function in t-1** | -236.491 | 0.0009 | -142.8018 | 8.45E-05 |

Notes: The results represent the following linear regressions using OLS estimation.

RV(t)=µ+β1*r(t-1)+β2*r(t)+β3*RV(t-1)+β4*I(t-1)*r(t-1) and

IV(t)=µ+β1*r(t-1)+β2*r(t)+β3*IV(t-1)+β4*I(t-1)*r(t-1).